\documentstyle[mathptm,pramana,epsf,floats]{ias}

\begin{document}
\title{Quark-Gluon Plasma: Status of Heavy Ion Physics}
\author{R.\ V.\ Gavai}
\address{Department of Theoretical Physics, Tata Institute of
Fundamental Research, Mumbai 400 005.} 

\abstract{
Lattice quantum chromodynamics (QCD), defined on a discrete space time 
lattice, leads to a spectacular non-perturbative prediction of a new
state of matter, called quark-gluon plasma (QGP), at sufficiently high 
temperatures or equivalently large energy densities. The experimental 
programs of CERN, Geneva and BNL, New York of relativistic heavy ion 
collisions are expected to produce such energy densities, thereby
providing us a chance to test the above prediction. After a brief
introduction of the necessary theoretical concepts, I will present a
critical review of the experimental results already obtained by the
various experiments in order to examine whether QGP has already been 
observed by them.}

\maketitle
\section{Introduction}

As is well known \cite{pdg}, the standard model of particle physics,
$SU(3)_c \times SU(2)_w \times U(1)_Y$ broken spontaneously to
$SU(3)_c \times U(1)_{em}$, has been tested with great precision at
LEP. All these tests rely heavily on the fact that the corresponding
coupling is weak and hence the usual weak coupling perturbation theory 
can be employed in deriving the required theoretical predictions. Since the
electromagnetic and weak couplings are indeed rather small in the
currently accessible energy range, $\alpha_{em} \simeq 7.3 \times
10^{-3}$ and $\alpha_w \simeq 3.4 \times 10^{-2}$, use of perturbation 
theory is not a serious limitation in the precision tests of the
electroweak theory. However, the strong interaction coupling,
$\alpha_s$, is (i) a strongly varying function of energy in the same
range, (ii) about 0.11 at the highest energy at which it has been
measured so far, and (iii) $\sim 1$ at a typical hadronic scale.
Therefore, testing the strongly interacting sector of the standard model 
using only perturbation theory is a major shortcoming of the precision
tests of the standard model.

Formulating quantum chromodynamics (QCD), which is an $SU(3)$ gauge
theory of quarks and gluons, on a discrete (Euclidean) space-time
lattice, as proposed by Wilson \cite{wil}, and simulating it
numerically as first shown by Creutz \cite{cre}, one can obtain
\cite{lat} several post-dictions of QCD in the non-perturbative domain 
of large $\alpha_s$. These include qualitative aspects, such as quark
confinement and chiral symmetry breaking, and quantitative details
such as hadron masses and their decay constants. While these agree
with the known experimental results within the sizeable theoretical
errors, it is fair to say that no serious experimental test of any
non-perturbative prediction of QCD has so far been made. Relativistic
heavy ion collisions offer a great window of opportunity to do
so. Application of lattice techniques to finite temperature QCD has
resulted in the prediction \cite{ftqcd} of a new state of matter,
called Quark-Gluon Plasma(QGP), at sufficiently high temperatures or energy 
densities. Chiral symmetry, broken spontaneously at zero temperature,
seems to be restored in this new phase characterised by a much larger
degrees of freedom characteristic of ``free'' quarks and
gluons. Nevertheless, the phase appears to be inherently
non-perturbative in the experimentally interesting range of $1 \leq
T/T_c \leq 4$-10, where $T_c \sim 150$ MeV is the transition
temperature at which the energy density varies most rapidly. The
energy density, $\epsilon$, in this range is 15-20\% smaller
\cite{ber} than the value of the corresponding ideal gas of quarks and 
gluons whereas a maximum of 3-5\% deviation is allowed for a weakly
interactive perturbative QGP. While the precise values for $\epsilon$, 
or $T_c$, as well as the nature of the phase transition (whether first 
order or second) depend on the number of light quark flavours, the
quoted values above being for 2 flavours of mass about 15 MeV, many
simulations with varying numbers of light flavours suggest that an
energy density greater than 1 GeV/fm$^3$ is needed to reach the QGP
phase. 

Collisions of heavy ions at very high energies can potentially produce 
regions with such large energy densities. Furthermore, since the
transverse size of such regions is given by the diameter of the
colliding nuclei, one can hope that these collisions will satisfy the
necessary thermodynamical criteria of large volume ($L \sim 2R_A \gg
\Lambda^{-1}_{QCD}$) and many produced particles. A crucial unanswered 
question is whether thermal equilibrium will be reached in these
collisions, and if yes, when it will be reached and how. Many
different attempts have been made, and are being made, to address
these issues. Here we will follow Bjorken's picture as it is most
widely used in the field. Bjorken argued \cite{bjo} that for
sufficiently high energies, $\sqrt{s} > 15 A$ GeV where $\sqrt{s}$ is
the total CMS energy of the two colliding nuclei of mass number $A$,
the nuclei bore through each other and leave behind a baryonless blob 
of produced particles in the center (around $y_{cm} = {1\over 2} \ln
~[(E + P_L) / (E - P_L)] \sim 0$). After an equilibration time
$\tau_0$, the energy density in the blob was estimated by Bjorken to
be 
\begin{equation}
\epsilon = {1\over {\cal A} \tau_0}\  \cdot \ {dE_T \over dy} \ ,
\label{eq:one}
\end{equation}
where the effective area ${\cal A} = \pi R^2_A = 3.94~A^{2/3}$ fm$^2$
and $dE_T/dy$ is the measured transverse energy per unit rapidity
round $y_{cm} \approx 0.0$.   

The Bjorken scenario for how the (thermally) equilibrated blob evolves 
is also the backbone of the analyses seeking to extract information
from the data on whether QGP did form in the heavy ion
collisions. According to this scenario, the hot blob cools by
expanding and the matter in it goes through various stages such as QGP,
mixed phase and a hadron gas, depending on the initial energy density
reached and the equation of state. A further rapid expansion of the
hadron gas leads to such large mean free paths for the hadrons that they
essentially decouple from each other.  If this freeze-out is
sufficiently fast, the free-streaming hadrons, $\pi, k, \cdots$
etc. will, however, retain the memory of the thermal state from
which they were born by having thermal momentum distributions. Thus
the information from observables related to light hadrons can tell us
about the temperature at this `thermal freeze-out' and the velocity of 
expansion. To get a glimpse at still earlier times, one has to turn to 
`harder' probes which typically involve a larger scale such
as masses of heavy quarkonia, as we will see below. 

The Bjorken scenario is for very high collision energies, when one expects 
to obtain a baryon-free region, and Eq. \ref{eq:one} is a valid
description.  The present collision energies may not be sufficient for it
to hold, i.e., there may be a lot of baryons deposited in the central
region of $y_{cm} \approx 0.0$.  A reliable analogue of Eq. \ref{eq:one}
is however not available in that case.  Note that even the theoretical 
estimate from lattice QCD above was for a baryonless case. 
In addition to temperature, one can also imagine increasing the baryon 
density of the strongly interacting matter or equivalently increasing
the baryonic chemical potential $\mu_B$ and obtain a baryon-rich plasma.  
In principle, one knows how to handle the case of a nonzero baryon density 
on the lattice but it has so far turned out to be difficult in practice.  
Usual lattice techniques fail for nonzero $\mu_B$ due to technical reasons
\cite{ftqcd} and attempts to overcome \cite{lat} these have not been
successful either.  No reliable lattice estimates are therefore available 
in that case.  Using models based on underlying symmetries, it has 
been recently argued \cite{ste} that the $T$ - $\mu_B$  phase diagram
of QCD with realistic quark mass spectrum should have a critical point 
at a nonzero $\mu_B$. The analysis of heavy-ion data by
Ref. \cite{ste} did not reveal any such critical point. By varying the 
energy of the colliding beam of heavy ions, one may hope to unearth
such a critical point. While upcoming experimental runs at CERN at
$\sqrt{s} \sim 9~A$ GeV will look for such a critical point, the model 
considerations above are inadequate to provide reliable information on 
the energy density of the QGP phase for nonzero $\mu_B$.  Thus a greater
theoretical effort is required to firm up the QCD prediction for the
energy density for nonzero $\mu_B$ and also to obtain the analogue of Eq.
\ref{eq:one}.    Of course, one can in stead go for higher energies to
test QCD, where one expects to obtain a baryon-free
region, making both the lattice estimate and Eq. \ref{eq:one} more
accurate descriptions.

\section{Results from CERN}

The experimental programs of high energy heavy ion collisions are
being pursued actively at present in Brookhaven National Laboratory
(BNL), New York and CERN, the European Laboratory for Particle
Physics, Geneva. $Au$-$Au$ collisions at $\sqrt{s} = 4.7~A$ GeV
$\simeq 0.92$ TeV have been studied at BNL while $Pb$-$Pb$ collisions
at $\sqrt{s} = 17.3~A$ GeV $\simeq 3.6$ TeV have been investigated at
SPS, CERN using beams of gold ions at 2.1 TeV/c and lead ions at 32.9 TeV/c 
respectively. Earlier sulphur beam at 6.4 TeV/c was used on sulphur
and uranium targets at SPS, CERN and those results form a benchmark over
which several aspects of $Pb$-$Pb$ collisions have been compared. I will
focus largely on the latter since they correspond to the highest
$\sqrt{s}$ used so far. Due to space restrictions, I will also have to 
restrict myself to highlights and I have to refer the reader for more
details to the proceedings of Quark Matter conferences \cite{qm}. 

\subsection{Initial Energy Density}

The NA49 experiment reported measurements on $dE_T/d\eta$ quite a
while ago \cite{na49} and reported $dE_T/dy \simeq 405$ GeV for $Pb$-$Pb$. 
Using a canonical guess of 1 fm for the formation time, one obtains from
Eq. \ref{eq:one}
\begin{equation}
\epsilon^{Pb-Pb}_{Bj} (1 {\rm fm}) = 2.94 \pm 0.3 {\rm GeV/fm}^3 \ ,
\label{eq:two}
\end{equation}
which is certainly above the characteristic QGP-phase values from
lattice QCD mentioned in sec. 1. Since appreciable numbers of baryons at 
$y_{cm} \sim 0$ have been observed at SPS, it is doubtful that the current 
energies are high enough for creating a baryon-free region assumed 
for Eq. \ref{eq:one}.  Nevertheless,  Eq. \ref{eq:one} has been
frequently used to estimate the energy densities achieved up to now.
One has to be cautious therefore and make sure that other independent
estimates are also similar and they do appear to be so. 

\subsection{Hadron Yields}

\begin{figure}[htbp]
\epsfxsize=8cm
\centerline{\epsfbox{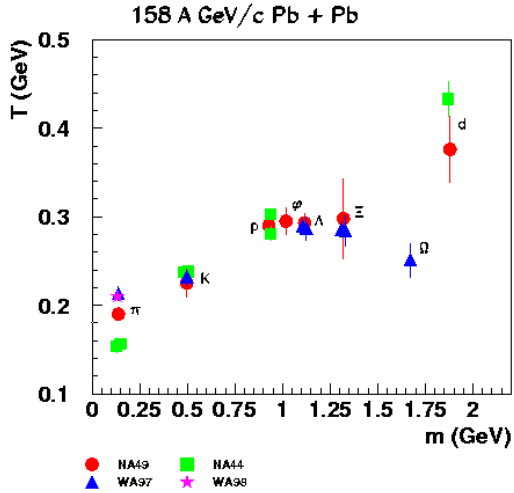}}
\caption{ $T_{\rm slope}$ for various particles as a function of mass.} 
\label{fig:one}
\end{figure}

Assuming that a thermal freeze-out is triggered by a rapid
expansion, one expects the momentum spectra of various hadrons to
reflect the freeze-out temperature, $T_{fo}$, which will be
blue-shifted by the collective expansion. For small transverse momenta, 
$p_T \ll m$, we expect the inverse slope of transverse mass
distribution, $d\sigma / d (m_T - m)$ with $m^2_T = p^2_T + m^2$,
to be given by $T_{\rm slope} = T_{fo} + {1\over 2} m \langle v_T
\rangle^2$. Thus one expects, $T_{\rm slope}$ to vary linearly with the
mass of the observed particle, $m$.  Fig. \ref{fig:one}
displays the dependence of $T_{\rm slope}$ on $m$ for various particles
produced in the $Pb$-$Pb$ collisions in the different CERN
experiments. A clear linear rise is evident for most of them except
the heavier strange particles for which freeze-out may be occurring
somewhat earlier. The $T_{fo}$ can be obtained from the intercept in
Fig. \ref{fig:one}, while the average collective velocity can be
obtained from the slope. 

\begin{figure}[htbp]
\epsfxsize=8cm
\centerline{\epsfbox{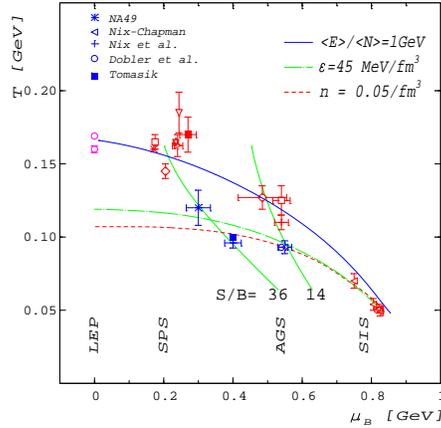}}
\caption{Chemical and thermal freeze-out points in the $(T, \mu_B)$ plane
for various experiments.  Taken from Ref. \protect \cite{clered}.} 
\label{fig:two}
\end{figure}

Strangeness changing -- chemical-- reactions are typically slower than
the elastic processes and hence are expected to freeze-out before the
thermal freeze-out. The temperature and chemical potential at this
freeze-out decides the particle yields of various types, provided these
yields are measured for the full 4$\pi$-integrated region; otherwise the
measurements will depend upon the details of the collective flow
mentioned above. Furthermore, taking ratios of such yields, one can
reduce the dependence on the collective dynamics even more. A simple
thermal model of free particles at a temperature $T$, volume $V$ and
chemical potential $\mu_B$ has been shown \cite{braun} to describe
beautifully 22 ratios of particle yields which vary by three
orders of magnitude, leading to $T^{chem}_{fo} \simeq 170$ MeV and
$\mu_{B,fo}^{chem} \simeq 270$ MeV. Fig. \ref{fig:two} displays the thermal
and chemical freeze-out points for the SPS $Pb$-$Pb$ collisions along
with those of other experiments. The figure is taken from Ref. \cite{clered}
where the references for the data are also given. A comment about
$\mu^{thermal}_{B,fo}$ may be in order, as we discussed above the
corresponding $T_{fo}$ only. Since chemical equilibrium is lost earlier,
it is strictly speaking not well defined. One simply adjusts 
$\mu_{B,fo}^{thermal}$ such that the particle ratios at 
$T^{therm}_{fo}$ agree with the observed values.

Since $T^{chem}_{fo}$ turns out to be very close to that expected for
the quark-hadron transition from lattice QCD, it is plausible that the
hadronic chemical equilibrium is a direct consequence of a pre-existing state 
of uncorrelated quarks and antiquarks and not due to hadronic
rescatterings/reactions, since there is not much time for the
latter. Hadron formation is then governed by the
composition of the earlier state in a statistical manner and an expansion 
later does not change their yields. Needless to say though, the
proximity of the two temperatures mentioned above is merely
indicative. Indeed such temperatures and chemical potentials could
still be reached via an expanding hadron gas as well.   One then would
expect though that the particle ratios will not reflect the underlying
quark symmetries described next.

\subsection{Excess Strangeness}

Since the early days of heavy ion collisions, when sulphur beams at
6.4 TeV were bombarded on sulphur targets, a global enhancement of
strangeness in these collisions has been observed relative to $e^+
e^-$ or $pp$ collisions. Defining \cite{Beca} a parameter $\lambda_s$
to count the strangeness, $\lambda_s = 2 \langle s + \bar s \rangle /
(\langle u + \bar u \rangle + \langle d + \bar d \rangle)_{\rm produced}$,
it was found that $\lambda^{AA}_s \simeq 2 \lambda^{PP}_s$. For both
$S$ (6.4 TeV) +$ Ag$ and $Pb$ (33 TeV) + $ Pb$, a similar factor of 2
enhancement was observed. This global enhancement together with the
picture of statistical hadron formation discussed above suggests an
interesting pattern for specific particles. 

\begin{figure}[htbp]
\epsfxsize=6cm
\centerline{\epsfbox{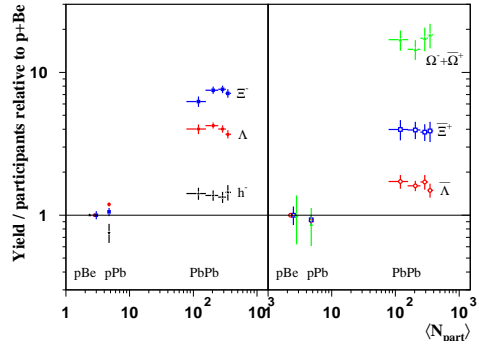}}
\caption{ Specific strangeness enhancement in Pb-Pb collisions seen by
the WA97 collaboration \protect \cite{wa97} as a function of number of
participants in the collisions.}
\label{fig:three}
\end{figure}
Fig. \ref{fig:three} shows the results from WA97 for specific
enhancements of $\Omega$, $\bar\Xi$ , $\bar\Lambda$ etc. The yield for
these particles per participant nucleon for $Pb$-$Pb$ relative to that 
in $p + Be$ is displayed as a function of average number of
participants. The number of participants is an indicator of how central
(or violent) the collision is; maximally central events are expected to
have largest energy density deposited. One sees the yields to be almost
independent of the centrality reached experimentally so far.  Assuming
now that the global strangeness enhancement by a factor of two indicates
the relative probability of finding a strange quark in $Pb$-$Pb$ vs
$p-Be$ to be twice, due perhaps to a formation of a QGP-state in the
latter which evaporates just before the observed chemical freeze-out,
one expects a factor of two enhancement for every extra strange quark
(or antiquark) in a hadron, explaining the pattern in Fig.
\ref{fig:three}. Note that the masses of these particles increase as
more strange quarks are added, $m_\Omega > m_\Xi > m_\Lambda$.  Their
production in a purely hadron rescattering/reactions scenario,
therefore, will be subject to increasingly higher thresholds, resulting
in an opposite pattern to that observed. Thus, the pattern in Fig.
\ref{fig:three} of specific enhancements of strangeness clearly {\it
points} to quark degrees of freedom. It will be interesting to see
whether this enhancement sets in smoothly or abruptly as the number of
participants vary.

\subsection{Excess Low Mass Dileptons}

An interesting anomaly, observed first in $S$ (6.4 TeV)+ $ Au$
collisions by NA45/CERES \cite{ceres} and confirmed for $Pb$ (33 TeV)
+$ Au$ collisions \cite{lenk}, is an enhancement of $M_{e^+e^-}$
spectrum in the region 250 MeV $< M_{e^+e^-} <$ 700 MeV. The data for
$p$ (450 GeV) +$ Au$ collisions in the entire range $0 \leq M_{e^+e^-} 
< $ 1500 MeV can be explained by including the contributions from all
known hadron decays of $\pi^0 , \eta , \omega , \rho , \phi , \eta'$
etc. within the acceptance of the detector. A similar exercise for
$S+Au$ or $Pb+Au$ reveals an enhancement in the low mass region
mentioned above, with the enhancement factor being 2.6 $\pm$ 0.5 (stat) 
$\pm$ 0.6 (syst). Various theoretical explanations have been offered 
\cite{rapp} to explain it. For us, it is interesting to note that 
thermal emission from an expanding fireball with parameters similar to 
those in sec. 2.2. can account for the excess, thus supporting to the picture
discussed there. 

\subsection{$J/\psi$ Suppression}

As remarked in the introduction above, one needs to employ `harder'
probes to explore the physics of the fireball at earlier times when
QGP may have existed. Production of $J/\psi$ is one such hard
probe. Since it is a tightly bound meson of charm and anticharm quarks, Matsui 
and Satz \cite{helmut} argued that color Debye screening of these heavy
quarks will prevent formation of $J/\psi$, if QGP is formed in the
heavy ion collisions. Due to a finite size and lifetime of the fireball, the
observable effect is expected to be a suppression in the production of
$J/\psi$. The NA38 and NA50 collaborations \cite{na50} measured
$J/\psi$ cross sections for a variety of collisions, starting from
$p+d$ to $Pb + Pb$ using the same muon spectrometer in the same
kinematic domain ($0 \leq y^{cm}_{\mu^+\mu^-} \leq 1$ and
$|\cos\Theta_{cs}| \leq 0.5$). While the systematic errors are thus
minimised, the lighter beams were necessarily of high energies;
$\sqrt{S_{NN}}$ thus varies from 17 GeV to 30 GeV. From the observed
dimuon spectrum, $dN/dM_{\mu^+\mu^-}$, obtained after subtracting the
background due to like-sign dimuons, the $J/\psi$ cross section was
obtained by a fit of the observed spectrum for $M_{\mu^+\mu^-} > 2.9$ GeV 
with 5 parameters: the number of events of Drell-Yan continuum, $J/\psi$, 
and $\psi'$ and the $J/\psi$ mass and width.

\begin{figure}[htbp]
\epsfxsize=7cm
\centerline{\epsfbox{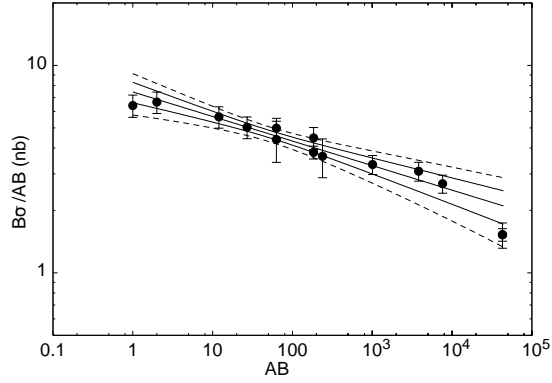}}
\caption{$J/\psi$ cross section times its branching fraction in to dimuons as 
a function of product of the mass numbers of target and projectile, AB. The 
data are from Ref. \protect \cite{na50} and the 1$\sigma$ (full lines) and
2$\sigma$(dashed lines) curves are from Ref. \protect \cite{us2}.} 
\label{fig:four}
\end{figure}

Comparing $\sigma^{DY}_{obs}$ with $\sigma^{DY}_{LO,th}$, where
isospin corrections were taken into account, the $K$-factor was found
to be universal in $pp$, $pA$ and $AB$ collisions: $\sigma^{DY}_{A\cdot
B} \propto A \cdot B$ for all of them, where $A$ and $B$ are the mass
numbers of the projectile and target respectively. Normalizing
$B_{\mu^+\mu^-} \sigma^{J/\psi}_{AB}$ by dividing by $A \cdot B$
therefore, where $B_{\mu^+\mu^-}$ is the branching fraction of
$J/\psi$ in to $\mu^+\mu^-$, one could expect QGP formation to be
signalled by a drop at some value of $A \cdot B$. Fig. \ref{fig:four}
shows the NA38 and NA50 results where one notices a gradual fall in with $A
\cdot B$ for {\it all} values. Note that some measurements have been re-scaled 
so that all are for the same energy in this figure. The decreasing cross
section for all values of $A \cdot B$, including small ones, is an indication 
of the presence of yet another mechanism for $J/\psi$-suppression in these 
collisions. Thus any suppression due to QGP will have to be over and above 
this `normal suppression'.

Production of heavy quarkonia is an old and mature area of
perturbative QCD. In particular, hadroproduction of $J/\psi$ has been
explained both in the colour evaporation model \cite{us1} and the
colour octet model \cite{gupsri} at $\sqrt{s}$ comparable to those in
Fig. \ref{fig:four}. So it is a natural question to ask whether 
the decrease in Fig. \ref{fig:four}
can be explained using pQCD. Unfortunately, sufficient information on
the nuclear structure functions, especially the gluonic ones, is not
available at present; assuming them to be independent of mass number
$A$ or $B$ is perhaps incorrect in view of the famous EMC-effect. 
Using the existing models of the EMC-effect, on the other hand, one finds 
hardly any decrease in the cross section in Fig. \ref{fig:four}. 
It is even likely that this lack of decrease of $B_{\mu^+\mu^-} 
\sigma^{J/\psi}_{AB} /AB$ with $AB$ is a generic feature independent of the
models of the EMC effect.  This is due to the fact that the dominant 
contribution to the cross section in Fig. \ref{fig:four} comes from the
so-called antishadowing region in $x$ which is hard to avoid for even the
gluons due to the momentum sum rule.  In view of the continuous decrease in
Fig. \ref{fig:four}, i.e. even for $p +$ light-$A$, where the radius of 
the target is only 2-4 times larger than that for the hadroproduction
\cite{us1,gupsri}, one has to ask whether a pQCD description of total
cross sections for $J/\psi$ is at all possible. It would be
interesting and desirable to thrash out this question by extensive
investigation of the nuclear glue and its impact on the $J/\psi$ cross
section.

The normal suppression in Fig. \ref{fig:four} has been explained
\cite{abs} as a final state interaction. The produced $J/\psi$-state or
its precursor can get absorbed in the nuclear matter (of the target and
beam). Treating $\sigma^{\psi N}_{abs}$ as a free parameter and using
the known nuclear profiles, one finds that a $\sigma_{abs} \sim$ 6.4 mb
can explain the linear fall in Fig. \ref{fig:four} quantitatively in
Glauber type models.  However, the $Pb$-$Pb$ data point seems to be off
this linear fall, and exhibits thus an `anomalous suppression'. One can
alternatively use an empirical $(AB)^\alpha$ fit to all points except
the $Pb$-$Pb$, which too will be linear on the scales of Fig.
\ref{fig:four} and the $Pb$-$Pb$ data point stands out again.

Unfortunately, the issue of how statistically significant this anomalous
suppression is gets affected by the crudeness of the theory described
above as well as by the assumptions needed to rescale some of the data
points. Ignoring these systematical theoretical errors, one finds the
anomalous suppression to be a 5$\sigma$ effect \cite{na50}, while
including them leads \cite{us2} to a conclusion that no anomalous
suppression exist at a 2$\sigma$ or 95\% confidence level, as shown by
the 2$\sigma$-band (enclosed by dashed lines) in Fig. \ref{fig:four}.

The NA50 collaboration also measures $J/\psi$-suppression as a function of 
the total produced transverse energy $E_T$. Using the same procedure as
outlined above, the number of $J/\psi$ events and the Drell-Yan events 
\footnote{Note that the usual Drell-Yan cross section in pQCD is
defined only by integrating all $E_T$.} in each $E_T$-bin are
determined. By taking their ratio, one obtains a less systematic error 
prone $R_{\rm expt} = B_{\mu^+\mu^-}
\sigma^{J/\psi}/\sigma^{DY}_{M_1-M_2}$ as a function of $E_T$, where
$M_1$-$M_2$ is the range of dimuon mass over which the Drell-Yan cross
section is integrated.  Using simple geometrical models, $E_T$ can be 
related to the impact parameter $b$ at which the two nuclei collide. 
Furthermore, any given $b(E_T)$ can be related to an average nuclear
path length $L$ which the produced $J/\psi$ (or its precursor) has to
traverse and which will determine the probability of its absorption in
nuclear matter.

\begin{figure}[htbp]
\epsfxsize=8cm
\epsfysize=6cm
\centerline{\epsfbox{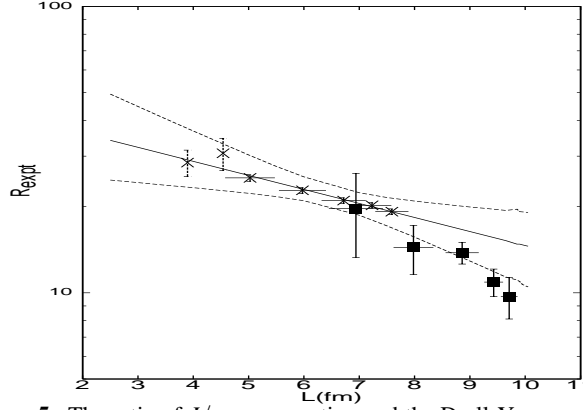}}
\caption{ The ratio of $J/\psi$ cross section and the Drell-Yan
cross section vs. $L$ in fm.  The crosses are NA38 data, shown along
with the straight line fit, a 4$\sigma$ band \protect \cite{us3} around it, 
and the NA50 data (squares) with 4$\sigma$ errors on them. The data are 
from Ref. \protect \cite{na50}.  } 
\label{fig:five}
\end{figure}

Fig. \ref{fig:five} shows $R_{\rm expt}$ as a function of $L$, as
determined by the NA50 collaboration, using $M_1$ = 2.9 and $M_2$ = 4.5. 
The normal nuclear suppression can be well approximated by $R_{\rm expt}
= A \cdot \exp (- \rho_{nucl} \cdot \sigma_{abs} \cdot L)$ or can be 
calculated more exactly in a Glauber model. The straight line in Fig.
\ref{fig:five} displays the fit for the light nuclei for $\rho_{nucl} =
0.17/$ fm$^3$ and $\sigma_{abs} \simeq 6.6$ mb. The low $L$ point
for $Pb$-$Pb$ collisions, corresponding to peripheral collisions, falls
on the fitted line while all the large $L$ points fall below it. Again,
one can ask for the statistical significance of this anomalous
behaviour. Since the fit above uses data from $E_T$-bins, or
equivalently $L$-bins, for lighter nuclei, there are again sizeable
errors on the theoretical prediction. For the 1995 data, which seem
broadly in agreement with the 1996 data and the 1998 data, it has been
estimated \cite{us3} that all the $Pb$-$Pb$ data points fall in a
4$\sigma$-band although they are all systematically below the
theoretical prediction, as shown in Fig. \ref{fig:five}.

It seems thus likely that an additional mechanism to suppress $J/\psi$ 
production in $Pb$-$Pb$ collisions, especially at large values of transverse
energy $E_T$, or large $L$, is needed over and above the normal
suppression caused by absorption in the surrounding nuclear matter. There 
have been several theoretical attempts to provide such a mechanism including, 
of course, invoking a possible a quark-hadron transition. A key non-QGP 
scenario invokes the possibility of destruction of the $J/\psi$ by the
so-called co-mover debris of the collisions. In a recent
\footnote{Although this came after my talk, I include it here for
completeness.} such work, it has been claimed
\cite{cap} that the {\it entire} NA50 $E_T$-spectrum of the $J/\psi$
cross section ratio $R$ can be explained using the co-mover
picture. In fact, the second shoulder in the $E_T$-behaviour observed
\cite{NewNa50} in the 1998 data and which could be due to QGP \cite{GuSa}, 
has been explained in Ref. \cite{cap} as due to fluctuations at the tail of 
the $E_T$-spectrum.  The difference between this mechanism and a QGP model 
will, therefore, most likely show up at the upcoming RHIC collider in BNL where 
$Au$ (19.7 TeV) + $Au$ (19.7 TeV) collisions will be studied this year
and the $E_T$ tail will extend much farther.

\section{Conclusions and Outlook}

An important non-perturbative prediction of (lattice) QCD is the
existence of a new phase of matter, Quark-Gluon Plasma, at
sufficiently high temperatures. Since the Standard Model has so far
been tested experimentally only in the weak coupling regime, it seems
desirable to confront this prediction with experiments. Collisions of
heavy ions at very high energy may be able to deposit the
required high energy density over a reasonable volume. The
experimental programs at BNL, New York and CERN, Geneva have by now
provided results for $Au$ on $Au$ and $Pb$ on $Pb$ at $\sqrt{s}
\simeq$ 0.9 TeV and 3.6 TeV (or $\sqrt{s}_{NN} \simeq$ 5 GeV and 17
GeV) respectively. The year 2000 should witness a factor of about 39
increase in the colliding CMS energy at BNL while LHC at 
CERN should achieve a $\sqrt{s}$ = 1150 TeV. The experiments so far have 
provided tantalizing hints of the new phase and therefore of the exciting
physics in the years ahead.

A fireball of QGP produced in these collisions cools by expanding
and converts into ordinary hadrons and leptons fairly quickly. Since
this makes a distinction of events with QGP formation from those without
it a very tough task, it seems prudent to look for a congruence of various
signatures in as many different ways of detecting QGP as
possible. Interestingly, the current results do indicate such a trend
of congruence from apparently unrelated measurements.

Soft hadron production data, such as $p_T$-distributions, two particle
(HBT) correlations and ratios of particle yields, can be interpreted
in terms of a chemical freeze-out, followed by a thermal
freeze-out. The freeze-out temperature for the former for the CERN SPS 
data turns out to be $\sim$ 170 MeV $\simeq T_c$ (quark-hadron
transition), suggesting that the hadrons were formed from an
uncorrelated QGP-like state. The global excess of strangeness by a
factor of two and the specific enhancement pattern of $\Omega$, $\bar
\Xi$ and $\bar \Lambda$, seen by the WA97 experiment, showing larger
enhancement for the heavier particles with more strange quarks is
consistent with this picture. In fact, the increasing thresholds make it
very difficult to explain this pattern in any conventional hadronic
picture. Thermal effects arising from $T \sim $ 120 MeV, the thermal
freeze-out temperature, can account for the observed excess of low mass
dielectron events of the NA45 experiment. 

Finally, anomalous $J/\psi$ suppression seen by the NA50 experiment
for $Pb$-$Pb$ collisions can be understood as arising out of a
deconfined quark-gluon plasma. Nevertheless, much more theoretical and 
experimental work will be needed to make a convincing case of
quark-gluon plasma formation in the heavy ion experiments since the
signals are still not spectacular in their statistical significance and
credible alternative explanations exist in many cases for the observed
results . Clearly, the commissioning of RHIC will be a big boost and will 
hopefully result in making a definitive case for quark-gluon plasma.


\begin{thebibliography}{99}
\bibitem{pdg}
See, e.g., Review of Particle Physics, Eur. Phy. J., {\bf C3}, 81 (1998).
\bibitem{wil} 
K. G. Wilson, Phys. Rev. {\bf D10}, 2445 (1974).
\bibitem{cre}
M. Creutz, Quarks, gluons and lattices, Cambridge Mongraphs (1983).
\bibitem{lat}
Proceedings of a) Lattice 98, Nucl. Phys. {\bf B(PS)73}, 1999, b)
Lattice 97, Nucl. Phys. {\bf B(PS)63}, 1998 etc.
\bibitem{ftqcd}
For a review, see, R. V. Gavai in ``Quantum fields on the computer'', Ed. 
M. Cruetz, World Scientific 1992, p. 51.
\bibitem{ber}
C. Bernard et al., Phys. Rev. {\bf D55}, 6861 (1997).
\bibitem{bjo}
J. D. Bjorken, Phys. Rev. {\bf D27}, 140 (1983).
\bibitem{ste}
M. Stephanov, K. Rajagopal, and E. Shuryak, Phys. Rev. {\bf D60}, 114028 (1999).
\bibitem{qm}
Proceedings of a) Quark Matter 99, Nucl. Phys. {\bf A661}, 1999, b)
Quark Matter 97, Nucl. Phys. {\bf A638}, 1998 etc.
\bibitem{na49}
NA49 Collaboration, T. Alber et al., Phys. Rev. Lett. {\bf 75}, 3814 (1995).
\bibitem{braun}
P. Braun-Munziger, I. Heppe and J. Stachel, Phys. Lett. {\bf B465}, 15 (1999).
\bibitem{clered}
J. Cleymans and K. Redlich, Phys. Rev. Lett. {\bf 81}, 5284 (1998).
\bibitem{Beca}
F. Becattini, M. Ga\'zdzicki and J. Sollfrank, Eur. Phys. J. {\bf C5},
143 (1998).
\bibitem{wa97}
WA97 Collaboration, E. Andersen et al, Phys. Lett. {\bf B449}, 401 (1999).
\bibitem{ceres}
CERES Collaboration, G. Agakichiev et al, Phys. Rev. Lett. {\bf 75}, 
1272 (1995).
\bibitem{lenk}
CERES Collaboration, G. Agakichiev et al, Phys. . Lett. {\bf B422}, 
405 (1998); B. Lenkeit in a) of Ref. \cite{qm}.
\bibitem{rapp}
R. Rapp in a) of Ref. \cite{qm}.
\bibitem{helmut}
T. Matsui and H. Satz, Phys. Lett. {\bf B178}, 416 (1986).
\bibitem{na50}
NA50 Collaboration, M. C. Abreu et al, Phys. Lett. {\bf B450}, 456 (1999);
{\it ibid}, {\bf B410}, 337 (1997).
\bibitem{us1}
R. V. Gavai, D. Kharzeev, H. Satz, G. A. Schuler, K. Sridhar and R.
Vogt, Int. J. Mod. Phys. {\bf A10}, 3043 (1995).
\bibitem{gupsri}
S. Gupta and K. Sridhar, Phys. Rev. {\bf D54}, 5545 (1996).
\bibitem{abs}
A. Capella, J. A. Casado, C. Pajares, A. V. Ramallo and J. Tran Thanh
Van, Phys. Lett. {\bf B206}, 354 (1988); C. Gerschel and J. H\"ufner,
Phys. Lett. {\bf B207}, 253 (1988); Z. Phys. {\bf C56}, 171 (1992). 
\bibitem{us2}
R. V. Gavai and S. Gupta, Phys. Lett. {\bf B408}, 397 (1997).
\bibitem{us3}
R. V. Gavai, Mod. Phys. Lett. {\bf A14}, 821 (1999).
\bibitem{cap}
A. Capella, E. G. Ferreiro and A. B. Kaidalov, {\tt hep-ph/0002300}.
\bibitem{NewNa50}
NA50 Collaboration, M. C. Abreu et al, CERN-EP-2000-013, Phys. Lett. 
{\bf B}, in press.
\bibitem{GuSa}
S. Gupta and H. Satz, Phys. Lett. {\bf B283}, 439 (1992); H. Satz, 
Nucl. Phys. {\bf A661}, 104c (1999).

\end{thebibliography}
\end{document}